\documentclass{ws-ijmpa}
\usepackage{url}
\usepackage[super,compress]{cite}
\usepackage{graphicx}
\usepackage{hyperref}
\hypersetup{colorlinks,urlcolor=black,citecolor=black,linkcolor=black,filecolor=black}
\usepackage{breakurl}
\begin{document}
\markboth{Cristina-Andreea Alexe}{Recent electroweak measurements from the CMS experiment}

%
\catchline{}{}{}{}{}
%

\title{Recent electroweak measurements from the CMS experiment}

\author{Cristina-Andreea Alexe \\ for the CMS Collaboration}

\address{Scuola Normale Superiore di Pisa \\
Piazza dei Cavalieri, 7, PI 56126, Pisa, Italy\\
Sezione di Pisa, Istituto Nazionale di Fisica Nucleare (INFN) \\
Largo Bruno Pontecorvo, 3, PI 56127, Pisa, Italy\\
cristinaandreea\_alexe@sns.it\\
cristina-andreea\_alexe@cern.ch}

\maketitle

\begin{history}
\received{Day Month Year}
\revised{Day Month Year}
\end{history}

\begin{abstract}
Recent measurements of electroweak phenomena from the Compact Muon Solenoid (CMS) experiment at the Large Hadron Collider are summarized. The standard model of particle physics was tested through highly precise determinations of its key electroweak parameters and through measurements of electroweak processes in proton-proton collisions at unprecedented center-of-mass energies of up to $13.6 \, \mathrm{TeV}$. The performance of the CMS experiment establishes its key role in the study of electroweak physics, with many measurements performed either for the first time or with the best precision at a proton-proton collider, in some cases reaching or even surpassing the precision of legacy results from lepton colliders. Recent electroweak results from the CMS experiment include: measurements of the W and Z bosons production cross sections; high-precision measurements of the forward-backward asymmetry in Drell-Yan production and of the effective leptonic electroweak mixing angle; measurements of tau lepton properties and of multiboson production and vector boson scattering rates.

\end{abstract}

\ccode{PACS numbers: 12.15.-y, 13.75.Cs, 13.85.-t}


\section{Introduction}	

The unified electroweak (EW) theory describes electromagnetic interactions mediated by a massless gauge boson, the photon, and weak interactions mediated by the $W^{\pm}$ and $Z$ gauge bosons, whose masses result from EW symmetry breaking via the Brout–Englert–Higgs mechanism. The gauge bosons interact among themselves and with other standard model (SM) particles through electric and weak charges. \cite{Wells:2017uxd}

At the Large Hadron Collider (LHC) the study of the EW sector is facilitated by the high production rate of gauge bosons in proton-proton collisions and by the excellent reconstruction of their various decay products with the Compact Muon Solenoid (CMS) detector. \cite{CMS:2024gzs} The study of leptonic final states is particularly powerful, as charged leptons stand out from the background dominated by quantum chromodynamics (QCD) processes.  

Theoretical predictions from perturbative calculations are available to a high precision for EW processes. This motivates the interpretation of EW measurement results from CMS in the context of the search for new physics (NP) in two categories: the precision and energy frontiers. For the former, very precise measurements of EW parameters such as the mass of the $W$ boson or the leptonic weak mixing angle are compared to theoretical predictions, and any deviation from the expectation could be attributed to NP processes that contribute to the SM process via virtual loops. On the other hand, analyses at high energies explore the possibility that deviations from gauge cancellations could lead to possibly large effects, detectable, for example, in multiboson production. Typically, these are effective field theory (EFT) interpretations or searches for anomalous couplings or axion-like particles (ALPs). The exploration of the energy frontier is additionally useful as it allows for stringent tests of QCD predictions and probes parton distribution functions (PDFs). Moreover, from an experimental point of view, it provides measurements of backgrounds for other analyses, e.g. in the Higgs boson sector and beyond the standard model (BSM) searches. 

CMS has a broad EW physics program, with many measurements performed either for the first time or with the best precision at a proton-proton collider, in some cases reaching or even surpassing the precision of legacy results from lepton colliders. This article covers the most recent EW results from CMS, among which: new measurements of the $W$ and $Z$ production cross sections in proton-proton collision of unprecedented center-of-mass energies of $13 \, \mathrm{TeV}$ and $13.6\, \mathrm{TeV}$ respectively; high-precision measurements of the forward-backward asymmetry in Drell-Yan production and of the effective leptonic electroweak mixing angle; measurements of tau lepton properties and of multiboson production and vector boson scattering rates.

\section{The CMS Detector}

The CMS apparatus is a general-purpose detector able to reconstruct electrons, muons, photons, and charged and neutral hadrons. Its defining feature is a $3.8 \, \mathrm{T}$ superconducting solenoid with an internal diameter of $6 \, \mathrm{m}$. Inside the solenoid there are an all-silicon inner tracker, a crystal electromagnetic (ECAL) and a brass-scintillator hadron (HCAL) calorimeters, each sectioned in a barrel and two endcap components. Forward calorimeters extend the pseudorapidity coverage of CMS, rendering it nearly hermetic. Gas-ionization muon detectors are installed in the flux-return yoke outside the solenoid. 
The LHC provides proton-proton collisions according to a schedule, relevant for this article being the period of Run 2 (2015-2018) at a center-of-mass energy of $13 \, \mathrm{TeV}$, and Run 3 (2022-ongoing) at $13.6\, \mathrm{TeV}$. A detailed description of the CMS detector together with the updated configuration for the LHC Run 3 are given in Refs. \citen{TheCMSCollaboration_2008},\citen{Hayrapetyan_2024}.

The performance on lepton detection and reconstruction at CMS is key to the success of its EW physics program. During Run 2, electrons and photons were reconstructed with an efficiency in data better than $95\%$ in the transverse energy $(E_\mathrm{T})$ range 10 to $500 \, \mathrm{GeV}$. Their energy was measured with an uncertainty on the scale of $<1\%$ in the $E_\mathrm{T}$ range 10 to $50 \, \mathrm{GeV}$, and within $2-3\%$ for higher energies, while the energy resolution for electrons from $Z$ decays is $2-5\%$. \cite{Sirunyan_2021} 

The efficiency for the reconstruction and identification of muons is $>96\%$. The muon momentum scale is corrected with an accuracy of up to $0.3\%$ and the resolution for muons up to $100 \, \mathrm{GeV}$ is between 1 and $3\%$ depending on pseudorapidity. \cite{Sirunyan_2018}

Neutrinos are invisible to the CMS detector, so they are inferred from the missing transverse momentum, obtained from the imbalance in the vector sum of objects reconstructed from the collision point.

\section{Precision Frontier}

\subsection{W and Z cross section measurements}

A measurement of the fiducial and total inclusive cross sections for $W$ and $Z$ boson production in proton-proton collisions at $5.02 \, \mathrm{TeV}$ and $13 \, \mathrm{TeV}$ was performed. \cite{cmscollaboration2024measurementinclusivecrosssections} Data samples obtained from dedicated runs with reduced instantaneous luminosity were used, so that a low average number of proton-proton interactions per bunch crossing (pileup) reduces the background contamination. The size of the data samples corresponds to integrated luminosities of $298 \, \pm \, 6 \, \mathrm{ pb^{-1}}$ at $5.02 \, \mathrm{TeV}$ and $206 \, \pm \, 5 \, \mathrm{ pb^{-1}}$ at $13 \, \mathrm{TeV}$. The analysis considers $W$ and $Z$ decays to electrons and muons and the cross sections are extracted from binned maximum likelihood fits of the transverse mass for the $W$ and dilepton invariant mass for the $Z$. The results for the total inclusive cross sections multiplied by the branching fractions at $5.02 \, \mathrm{TeV}$ are $ \sigma(\mathrm{pp\xrightarrow{}}W\mathrm{+X})\mathcal{B}(W\xrightarrow{}l\nu) = 7300 \pm10\mathrm{(stat)} \pm60\mathrm{(syst)} \pm140\mathrm{(lumi)} \, \mathrm{pb} $ and $ \sigma(\mathrm{pp\xrightarrow{}}Z\mathrm{+X})\mathcal{B}(Z\xrightarrow{}l^+l^-) = 669 \pm2\mathrm{(stat)} \pm6\mathrm{(syst)} \pm13\mathrm{(lumi)} \, \mathrm{pb} $, and at $13 \, \mathrm{TeV}$ they are $ 20480 \pm10\mathrm{(stat)} \pm170\mathrm{(syst)} \pm470\mathrm{(lumi)} \, \mathrm{pb} $ and $ 1952 \pm4\mathrm{(stat)} \pm18\mathrm{(syst)} \pm45\mathrm{(lumi)} \, \mathrm{pb} $ respectively, where “stat”, “syst” and “lumi” denote, respectively, the statistical uncertainty due to the sample size, the systematic uncertainty due to experimental and theoretical effects, and the uncertainty due to the measurement of the proton beam luminosity. The leading uncertainty comes from the luminosity measurement. Ratios of cross sections of $W^+$ and $W^-$ and of inclusive $W$ and $Z$ production are convenient to measure, as the luminosity and other systematic uncertainties cancel, allowing for a better comparison with theoretical predictions. Measurements of production cross sections are important tests of QCD through the probing of proton PDFs. The results of this analysis are in agreement with the theoretical predictions at next-to-next-to leading order (NNLO) in perturbative QCD. Compared to previous similar measurements at the LHC, the presented analysis has reduced systematic uncertainties. The fiducial cross section for $W$ and $Z$ production at $5.02 \, \mathrm{TeV}$ reached a precision of less than $2\%$. Additionally, ratios of cross sections at two center-of-mass energies are measured.

A more recent measurement of the $Z$ production cross section in proton-proton collisions at $13.6 \, \mathrm{TeV}$ was performed, based on $Z$ decays to two muons. \cite{CMS-PAS-SMP-22-017} The resulting total cross section is $ \sigma(\mathrm{pp\xrightarrow{}}Z\mathrm{+X})\mathcal{B}(Z\xrightarrow{}\mu^+\mu^-) = 2.010 \pm0.001\mathrm{(stat)} \pm0.018\mathrm{(syst)} \pm0.046\mathrm{(lumi)} \pm0.007\mathrm{(theo)} \, \mathrm{nb} $, where "theo" denotes uncertainties due to theory modeling. The uncertainty is dominated by systematics and the result is in agreement with the theoretical prediction at NNLO in perturbative QCD.  

\subsection{W and Z boson decays}

Measurements of the branching fraction for $Z$ decays to four leptons, considering electrons and muons, were performed at $8 \, \mathrm{TeV}$ and $13 \, \mathrm{TeV}$.\cite{CMS-PAS-SMP-19-007} The measured value of the inclusive branching fraction for all four-lepton decay modes $ \mathcal{B}(Z\xrightarrow{}4l) = (4.67 \pm0.11\mathrm{(stat)} \pm0.10\mathrm{(syst)}) \times 10^{-6} $ has a precision of $\sim \,$3\%, better than previous results by the CMS and ATLAS experiments at the LHC. The reported branching fractions and differential decay rates are consistent with the SM predictions. Charge conjugation and parity (CP) invariance violations are probed through measurements of triple-product asymmetries, consistent with the SM. Improved limits are obtained on NP in the form of a scalar or vector boson mediating the $Z\xrightarrow{}4l$ process.

The first search for the decay $Z\xrightarrow{}\tau\tau\mu\mu$ at the LHC was performed at $8 \, \mathrm{TeV}$ and $13 \, \mathrm{TeV}$. \cite{PhysRevLett.133.161805} No excess over the SM background was observed and an upper limit at 95\% confidence level (CL) was placed on the ratio of the $Z\xrightarrow{}\tau\tau\mu\mu$ to $Z\xrightarrow{}4\mu$ branching fractions equivalent to 6.9 times the SM expectation of $0.90 \pm 0.02$. Additionally, the first constraints on all flavor-conserving 4-lepton Wilson coefficients involving 2$\mu$ and 2$\tau$ were given.

The most precise measurement to date of the $W$ boson hadronic decay branching fraction ratio $ \mathrm{R}_c^W = \mathcal{B}(W\xrightarrow{}cq)/\mathcal{B}(W\xrightarrow{}q\bar{q}')$ was performed on proton-proton collisions at $13 \, \mathrm{TeV}$. \cite{cmscollaboration2024measurementwbosondecay} The result $ \mathrm{R}_c^W = 0.489 \, \pm \, 0.020 $ is consistent with the SM and provides an increase in precision by a factor of two to the world average. The sum of squared elements in the second row of the Cabibbo-Kobayashi-Maskawa (CKM) matrix and the CKM matrix element $\left| V_\mathrm{CS} \right|$ were determined to be $0.970 \pm 0.041$ and $0.959 \pm 0.021$ respectively.  

\subsection{The effective leptonic weak mixing angle}

A measurement of the forward-backward asymmetry ($A_\mathrm{FB}$) in Drell-Yan production and of the effective leptonic electroweak mixing angle ($\theta^l_\mathrm{eff}$) in proton-proton collisions at $13 \, \mathrm{TeV}$ was performed using a full Run 2 data set of 138 $\mathrm{fb^{-1}}$. \cite{cmscollaboration2024measurementdrellyanforwardbackwardasymmetry} This represents an important test of the SM, through a comparison with the precise theoretical prediction of $\sin^2\theta^l_\mathrm{eff} = 0.23155 \pm 0.00004$. \cite{10.1093/ptep/ptac097}

The electroweak mixing angle $\theta_W $ relates the masses of the $W$ and $Z$ bosons as $\sin^2\theta_W = 1 - m^2_W/m^2_Z$, while the effective corresponding quantity absorbs EW radiative corrections to leading order (LO) relations of Z bosons couplings to fermions. The parameter $\sin^2\theta^l_\mathrm{eff}$ is accessed from the angular distribution of final state leptons from Z decays to muon or electron pairs in the Collins-Soper (CS) reference frame \cite{PhysRevD.16.2219}, given at LO and for given dilepton invariant mass by

\begin{equation}
\frac{d\sigma}{d\mathrm{(cos\,\theta_\mathrm{CS})}} \propto 1 + \mathrm{cos^2\,\theta_\mathrm{CS}} + A_4\mathrm{cos\,\theta_\mathrm{CS}},
\label{diseqn}
\end{equation} 

where $\theta_\mathrm{CS}$ is the angle between the negative lepton and the axis bisecting the angle between the direction of the quark and the reversed direction of the antiquark that produced the $Z$ boson, and $A_4$ is an angular coefficient.

The forward-backward asymmetry is defined as

\begin{equation}
A_\mathrm{FB} = 3/8 A_4 = \frac{\sigma_\mathrm{F}-\sigma_\mathrm{B}}{\sigma_\mathrm{F}+\sigma_\mathrm{B}},
\label{diseqn}
\end{equation} 

where $\sigma_\mathrm{F}$ and $\sigma_\mathrm{B}$ are the cross sections in the forward ($\cos \theta_\mathrm{CS} > 0 $) and backward ($\cos \theta_\mathrm{CS} < 0 $) hemispheres, respectively. $A_\mathrm{FB}$ is mostly sensitive to $\sin \theta^l_{eff}$. \cite{CMS:2018ktx}

In this analysis, the determination of $\sin^2\theta^l_\mathrm{eff}$ is performed in two ways. The first one follows the method of the previous CMS analysis at $8 \, \mathrm{TeV}$ \cite{CMS:2018ktx}, using the directly measured forward-backward angular-weighted asymmetry, approximately equal to $A_\mathrm{FB}$, with weights that are functions of the rapidity and $\cos \theta_\mathrm{CS}$ of the dilepton. The second method extracts $\sin^2\theta^l_\mathrm{eff}$ from template fits to $A_\mathrm{FB}$ as a function of dilepton mass and rapidity before final-state radiation. While the first method benefits from the cancellation of experimental systematic uncertainties and the large data sample, resulting in improved final uncertainties with respect to Ref. \citen{CMS:2018ktx}, the second, more complex method leads to a reduced dependence of the result on theory and PDFs, with benefits in the interpretation and use of the result.

Since the analysis is based on the measurement of lepton angles, which depends on the rapidity of the dilepton, the description of valence quarks in the PDFs becomes important, making the measurement of $A_\mathrm{FB}$ sensitive to the PDF choice. The chosen default PDF set is CT18Z. However, compatible results within their respective uncertainties are also obtained with other PDF sets when profiling the PDF uncertainty in the likelihood fit.

The result obtained through the first method has smaller uncertainties compared to the one obtained with the second method, and is given by $\sin^2\theta^l_\mathrm{eff} = 0.23157 \, \pm \, 0.00031 $, which represents the best measurement at a hadron collider. This precise result weights in the ambiguity generated by the discrepancy between lepton and b-quark results at the Large Electron–Positron Collider (LEP).

\subsection{Tau lepton properties}

CMS has achieved the first observation at a proton-proton collider of tau lepton pair production via photon-photon fusion $\gamma\gamma\xrightarrow{}\tau\tau$ with a significance of 5.3 standard deviations, using the full Run 2 data set at $13 \, \mathrm{TeV}$. The observation enables the setting of constraints on the contribution that new physics could have to the anomalous magnetic moment $  \mathrm{a}_{\tau} = 0.0009^{+0.0032}_{-0.0031} $ and electric dipole moment $ \left|  \mathrm{d}_{\tau} \right| < 2.9 \times 10^{-17} e \, \mathrm{cm}$ (95$\%$ CL) of the tau lepton. This represents the most stringent limit on the tau magnetic moment to date, improving previous constraints by nearly an order of magnitude.

Events are selected by requiring tau leptons to be produced back-to-back in the azimuthal direction, and by imposing an upper threshold on the number of charged hadrons associated with their production vertex. The tau leptons were reconstructed from their leptonic and hadronic decay modes. The fiducial cross section of $\gamma\gamma\xrightarrow{}\tau\tau$ was measured to be $\mathrm{\sigma_{fid}} = 12.4^{+3.8}_{-3.1} \, \mathrm{fb} $. \cite{CMS_Collaboration_2024}

The polarization of tau leptons was measured using $Z\xrightarrow{}\tau\tau$ events in Run 2 data at $13 \, \mathrm{TeV}$ and it is given by 

\begin{equation}
\mathcal{P}_{\tau} = \frac{\sigma_+-\sigma_-}{\sigma_++\sigma_-},
\label{diseqn}
\end{equation} 

where $\sigma_+$ and $\sigma_-$ are the production cross sections for tau leptons with positive and negative helicities, respectively. The polarization was determined to be $\mathcal{P}_{\tau} = -0.144 \, \pm \, 0.015 $, which is the most precise measurement at a hadron collider, in agreement with results from the SLD experiment at the Stanford Linear Accelerator Center (SLAC) and LEP experiments.\cite{CMS:2023mgq} The polarization constrains the effective couplings of tau leptons to the $Z$ boson and leads to the determination of  $\sin^2\theta^l_\mathrm{eff} = 0.2319 \, \pm \, 0.0019$, therefore with a precision of 0.8\% and independently of the Z production. 

\section{Energy Frontier}

\subsection{Multiboson production}

In CMS, couplings involving three or four gauge bosons are studied from proton-proton events with diboson and triboson final states respectively. Recent results from Run 3 data at $13.6 \, \mathrm{TeV}$ include the production of $W^+W^-$, with $W$ decays to electrons or muons, with a measured cross section $\sigma_{WW} = 125.7 \, \pm \, 5.6 \, \mathrm{pb}$ \cite{CMS:2024hey} and the production of $WZ$ \cite{CMS:2024ild}, one of the cleanest diboson channels at the LHC with three leptons in the final state, with a measured cross section $\sigma_{WZ} = \mathrm{55.2 \pm 1.2(stat) \pm 1.2(syst) \pm 0.8(lumi) \pm 0.1(theo)} \, \mathrm{pb}$. Both results are in agreement with QCD predictions at NNLO and EW predictions at next-to leading order (NLO). Another diboson process studied from Run 2 data at $13 \, \mathrm{TeV}$ is the production of a $Z$ boson (decaying to two neutrinos) and of a photon, $Z(\nu\nu)\gamma$, allowing for the determination of the most stringent limits on the anomalous neutral triple gauge coupling from CMS. The selection of relevant events relies on measuring photons with high transverse momentum, which at CMS can reach up to over 1 TeV. \cite{CMS-PAS-SMP-22-009}

Recent results on triboson production are from Run 2 at $13 \, \mathrm{TeV}$ and include the first observation of $WW\gamma$ production at a proton-proton collider, with a significance of 5.6 standard deviations. \cite{CMS:2023rcv} A fully leptonic final state is considered and the resulting cross section $\sigma_{WW\gamma} = 5.9 \pm 1.3 \, \mathrm{fb}$ is in agreement with the QCD prediction at NLO. Additionally, the study of $WZ\gamma$ production with fully leptonic final states allowed the setting of limits on the anomalous quartic gauge coupling and on photophobic axion-like particle models. The measured fiducial cross section $ \sigma^{WZ\gamma}_\mathrm{fid} = 5.48 \, \pm \, 1.11 \, \mathrm{fb} $ is in agreement with the QCD prediction at NLO. \cite{CMS-PAS-SMP-22-018}

\subsection{Vector boson scattering}

CMS achieved the first study of same-sign $W$ boson scattering with a hadronic tau in the decay channel. The employment of deep neural network algorithms allowed the discrimination of the signal from the main backgrounds. The measured cross section is $1.44^{+0.63}_{-0.56}$ times the SM prediction. This process has access to the quartic electroweak coupling, and the results were interpreted in an EFT analysis. No deviations from the SM expectations were found. This represents the first study of EFT operators with different dimensions. \cite{CMS:2024jvu}

\section{Conclusion}

The CMS experiment has a large and diversified electroweak physics program, with a series of significant results in the precision and energy frontiers. Recently, CMS performed tests of the standard model with proton-proton collisions at center-of-mass energies of 8, 13, and $13.6 \, \mathrm{TeV}$ by measuring production cross sections of EW processes or by measuring key EW parameters. Notable examples are the best measurement of $\sin^2\theta^l_\mathrm{eff}$ at a hadron collider and measurements of tau lepton properties. These measurements are reaching or even surpassing the precision of legacy results from lepton colliders. Tests of the EW theory at the highest energies are performed in multiboson measurements, a recent notable result being the first observation of $WW\gamma$ in proton-proton collisions.

The performance of the CMS experiment is achieved by successfully exploiting the large amounts of collisions provided by the LHC, in spite of the challenging high pileup environment, many results being limited by systematic uncertainties. CMS analyses have been constantly improving the precision of EW measurements, and no deviations from the standard model have been observed. Thus, the need for theoretical predictions of higher precision is starting to appear, such as QCD predictions at next-to-next-to-next-to leading order.

\section*{Acknowledgments}

C.-A. Alexe acknowledges financial support from the European Research Council (ERC) under the European Union’s Horizon 2020 research and innovation program (Grant agreement N. 10100120).

\bibliographystyle{ws-ijmpa}
\bibliography{sample}

\end{document}